\documentclass[11pt]{article}
\usepackage{epsfig}
\usepackage{amsmath}
\usepackage{amssymb}
\usepackage{amsfonts}
\usepackage{mathrsfs}
\newcommand{\bc}{\begin{center}}
\newcommand{\ec}{\end{center}}
\newcommand{\bt}{\begin{tabular}}
\newcommand{\et}{\end{tabular}}
\newcommand{\be}{\begin{equation}}
\newcommand{\ee}{\end{equation}}
\newcommand{\bea}{\begin{eqnarray}}
\newcommand{\eea}{\end{eqnarray}}
\newcommand{\bfig}{\begin{figure}}
\newcommand{\efig}{\end{figure}}
\def\be{\begin{equation}}
\def\ee{\end{equation}}
\def\bea{\begin{eqnarray}}
\def\eea{\end{eqnarray}}

\begin{document}

\bc
{\Large \bf HEAVY SCALAR DARK MATTER}\\
\vspace{0.5cm}
{\large FU-SIN LING}\\
\vspace{0.5cm}
{\it Service de Physique Th\'eorique \\
Universit\'e Libre de Bruxelles}
\ec

\begin{abstract}
Scalar fields with $SU(2)_L$ quantum numbers provide several viable Dark Matter candidates around the TeV scale.
Thanks to scalar quartic interactions, the observed relic density can be obtained for a large range of masses,
which we determine. In turn, the constraints on the scalar couplings lead to testable predictions in future
direct and indirect searches.
\end{abstract}

\section{Minimal models of Dark Matter}

The WIMP miracle has often been put forward as an argument in favor of theories like Supersymmetry.
Stripped to the core, it states that a Dark Matter (DM) candidate with a mass around 1~TeV and a typical weak
interaction annihilation cross-section should have a relic density of order one (in units of the critical density).
The possibility of a DM candidate in Supersymmetry is therefore a free gift.

One could instead build a minimal model of Dark Matter, and would therefore naturally be led to consider
extra fields with $SU(2)_L$ quantum numbers. By minimality, we mean that only one extra field is added to the Standard Model,
so that the number of new parameters is limited. 
Such a criterion is important with respect to the predictivity and the testability of the model. 
A parity symmetry under which the extra field is odd while all Standard Model 
fields are even should also be added in order to guarantee the stability of DM.

For a fermionic field, only gauge interactions are allowed by renormalizability. Therefore, the observed DM relic abundance
can only be obtained for a specific value of the DM mass, which depends on the dimension of the $SU(2)_L$ multiplet containing
the extra field~\cite{Cirelli:2005uq}.
For a scalar field however, renormalizability allows for scalar quartic interactions on top of gauge interactions.
Therefore, the most general scalar case, although minimal, should include this possibility.
This freedom allows to have viable DM candidates within an extended mass range around the TeV scale.
For a precise determination of this mass range (for each multiplet), perturbativity and stability constraints have to be taken into account.
Moreover, the epoch of the electroweak phase transition also plays an important role in the case of a doublet.
Finally, for higher multiplets, non-perturbative enhancements known as Sommerfeld effects are far from negligible.
We refer the reader to Ref.~\cite{ourpaper} (and references therein) for technical details.

Here we will only describe general features of scalar multiplet models and stress the role of the scalar interactions.
Let us denote by $H_1$ the usual Brout-Englert-Higgs scalar $SU(2)_L$ doublet of the Standard Model (SM).
We add an extra scalar multiplet $H_n$ to the SM, odd under some $Z_2$ symmetry, and with $n$ being its dimension. 
Two distinct cases have to be considered, namely the doublet case and the higher multiplet case.
Indeed, if the multiplet is a doublet, bilinears $H_1^\dagger H_2$ can be constructed, leading to a different potential.

For the doublet case, the most general renormalizable potential is
\begin{equation}
\label{potential}
\begin{split}
V(H_1,H_2) &= \mu_1^2 \vert H_1\vert^2 + \mu_2^2 \vert H_2\vert^2  + \lambda_1 \vert H_1\vert^4
+ \lambda_2 \vert H_2\vert^4 \\
&+ \lambda_3 \vert H_1\vert^2 \vert H_2 \vert^2 + \lambda_4 \vert H_1^\dagger H_2\vert^2
+ {\lambda_5\over 2} \left[(H_1^\dagger H_2)^2 + h.c.\right]~.
\end{split}
\end{equation}
After the electroweak symmetry breaking, $H_1$ develops its vev, $v_0 = -\mu_1^2/\lambda_1 \simeq 246$ GeV, 
which leads to mass splittings between charged and neutral components of $H_2=(H^+ \quad (H_0+iA_0)/\sqrt{2})^T$.
We have
\bea
m_h^2 &=& 2 \lambda_1 v_0^2~,\cr
m_{H_0}^2 &=& \mu_2^2 +  \lambda_{H_0} v_0^2~,\cr
m_{A_0}^2 &=& \mu_2^2 +  \lambda_{A_0}  v_0^2~,\cr
m_{H^+}^2 &=& \mu_2^2 + \lambda_{H_c} v_0^2~,
\label{masses}
\eea
with $\lambda_{H_c} \equiv \lambda_3/2$ and $\lambda_{H_0,A_0} \equiv (\lambda_3 + \lambda_4 \pm \lambda_5)/2$.
In particular, the last term of the potential Eq.~(\ref{potential}), when present, generates a mass splitting 
between the neutral components of $H_2$. To have a viable DM candidate, such a mass splitting is necessary to avoid too large elastic scattering
cross-sections in direct detection experiments through a vector coupling to the Z boson.
As we will see, for higher multiplets, such a mass splitting cannot be generated at this level of minimality in the lagrangian,
therefore excluding all the models with a non-zero hypercharge.
The combinations $\lambda_{H_0}$, $\lambda_{A_0}$ and $\lambda_{H_c}$ that appear in the mass spectrum also play the role 
of scalar quartic couplings in the potential. Therefore, the phenomenology of the so called Inert Doublet Model
is completely determined by these couplings and the mass of the DM candidate, conventionally chosen as $H_0$.  

For higher multiplets, the most general renormalizable potential is
\begin{equation}
\label{eq:potentialmultiplet}
\begin{split}
V(H_n,H_1) = V_1(H_1) &+ \mu^2 H_n^\dagger H_n +
\frac{\lambda_2}{2} \left(H_n^\dagger H_n\right)^2 + \lambda_3
\left(H_1^\dagger H_1\right) \left(H_n^\dagger
H_n\right) \\
&+ \frac{\lambda_4}{2} \left(H_n^\dagger \tau_a^{(n)} H_n\right)^2
+ \lambda_5 \left(H_1^\dagger \tau_a^{(2)} H_1\right)
\left(H_n^\dagger \tau_a^{(n)} H_n\right)~,
\end{split}
\end{equation}
where a sum over $a$ is implicit in the last two terms, and $\tau_a^{(n)}$ are the $SU(2)$ generators for the representation of dimension $n$.
As announced, no term in the potential Eq.~(\ref{eq:potentialmultiplet}) can generate a mass splitting between the real and the imaginary parts
of the neutral component of $H_n$. If $Y \neq 0$, the DM candidate would couple to the Z boson, with elastic scatterings orders of magnitude
above current detection limits~\cite{Ahmed:2008eu,Alner:2007ja}. For $Y=0$, the multiplet can still be real or complex. In Ref.~\cite{ourpaper}, we argue that
phenomenologically viable complex multiplet models are very similar to real multiplet models, except for the doubling of the number of fields.

For real multiplets, bilinears with $SU(2)$ generators identically vanish, so that the potential Eq.~(\ref{eq:potentialmultiplet}) reduces to
\begin{equation}
\label{finalotentialmultiplet}
V(H_n,H_1) = V_1(H_1)+ \mu^2 H_n^\dagger H_n +
\frac{\lambda_2}{2} \left(H_n^\dagger H_n\right)^2 + \lambda_3
\left(H_1^\dagger H_1\right) \left(H_n^\dagger H_n\right) \quad .
\end{equation}
At tree-level, the mass spectrum is degenerate, with a common mass
\be
\label{eq:massmultiplet} 
m^2_0 = \mu^2 + \frac{\lambda_3 v_0^2}{2} \quad . 
\ee
This degeneracy is lifted by radiative corrections which increase the mass of charged components by a few hundreds MeV~\cite{Cirelli:2005uq}.
An important consequence of this analysis is that the phenomenology of higher multiplet models ({\it w.r.t.} DM) is 
completely determined by only one scalar quartic coupling, namely $\lambda_3$, whereas three such couplings are present in the doublet case.

\section{Scalar vs. gauge interactions}

When scalar quartic couplings are absent, all DM states are degenerate at tree-level. The pure gauge annihilation cross-section is a function
of the DM mass only. In a standard thermal freeze-out scenario, the relic abundance of the DM candidate roughly scales as 
the inverse thermal average of the annihilation cross-section $\Omega_{DM} \propto \langle \sigma v \rangle^{-1}$.
The latest five-year WMAP combined result on the DM density $\Omega_{\rm DM} h^2 = 0.1131 \pm 0.0034$~\cite{Hinshaw:2008kr}
therefore fixes the mass of the DM candidate. 
These threshold values for all candidates of phenomenological interest are given in Table~\ref{scatab}.
\begin{table}
\small
\begin{center}
\begin{tabular}{lccccc}
\hline \hline
Models & $\lambda_3=0 $ & $\lambda_3=2\pi $ & $\lambda_3=4\pi $ & $\lambda_3=0$ (SE) & $\lambda_3=4\pi$ (SE) \\
\hline
Inert Doublet & $0.534 \pm 0.0085$ & 22.5 & 46 & 0.55 & 47 \\
Real Triplet & $1.826 \pm 0.028$ & $11.1$ & $21.9$ & 2.3 & 28.1\\
Real Quintuplet & $4.642 \pm 0.072$ & $9.6$ & $17.4$ & 9.4 & 35.7\\
Real Septuplet & $7.935 \pm 0.12$ & $10.6$ & $16.1$ & 22.4 & 46.3\\
\hline \hline
\end{tabular}
\end{center}
\caption{{\it Threshold masses (in TeV) without or with Sommerfeld effect (SE) for scalar multiplet models, as 
determined by the WMAP constraint, the errors quoted correspond to a 1$\sigma$ variation of the relic density.
The large mass range of the DM candidate is shown by the indicative values for $\lambda_3=2\pi$ and $4\pi$.}}
\label{scatab}
\end{table}
Without scalar interactions, the scalar DM candidate has a mass in the TeV range, and annihilates mainly into
$W$, $Z$ bosons and photons. Coannihilation channels into fermions pairs and $Z h$ (or $W h$) are non negligible. 

With the scalar quartic couplings present, the doublet case has to be distinguished from the higher multiplet case
because mass splittings between the doublet components are generated. Therefore, in the doublet case, it is possible to 
suppress the coannihilation channels even for a light DM mass. It has been shown that the Inert Doublet Model can
give rise to the correct relic density in three possible regimes : the low-mass, the middle-mass and the high-mass regimes.
Here we will focus on the high-mass regime only. For higher multiplet models however, the multiplet components stay degenerate
even when scalar quartic couplings are switched on. As a consequence, higher multiplet models are compatible with the relic density
constraint only in the high-mass regime.

In the high-mass regime, it turns out that the total annihilation cross-section relevant for the calculation of the relic density
can only increase when scalar quartic couplings are turned on. For higher multiplet models, this is because the Higgs pair channel 
is opened while the pure gauge channels are not modified. For the doublet case, the analysis is more subtle and is deeply connected
to gauge invariance. Indeed pure gauge annihilations for $\lambda = 0$ mainly produce transverse modes of gauge bosons.
Any annihilation amplitude into a pair of longitudinal modes of $W$ (or $Z$) is suppressed by a factor $m_{H_0}^2/m_W^2$. 
This residual amplitude is the result of a cancellation between various amplitudes (point-like, $t$ and $u$-channels). 
In the high-mass regime, this residual amplitude is completely negligible compared to the transverse amplitude.
When scalar quartic couplings are switched on, the cancellation is lost since the doublet partners of the DM candidate can have a higher mass.
Therefore the annihilation amplitude into gauge bosons picks up a longitudinal contribution proportional to a mass splitting 
between the odd fields. As cross-sections into transverse and longitudinal modes add up quadratically, it is clear that the 
total annihilation cross-section can only increase when scalar quartic couplings are switched on.
The scalar contribution to the cross-section becomes comparable to the pure gauge one for $\lambda \simeq 1$.

\section{Relic abundance}

The discussion on how scalar quartic couplings increase the annihilation cross-section shows that scalar multiplet models can fulfill
the WMAP abundance requirement for any DM mass above the threshold values in Table~\ref{scatab}.

In the doublet case, this constraint translates into an upper bound for each scalar quartic coupling and for the mass splittings,
as shown on Fig.~\ref{highlambdamax}. 
\bfig
\bc
\bt{cc}
\includegraphics[width=0.4\textwidth]{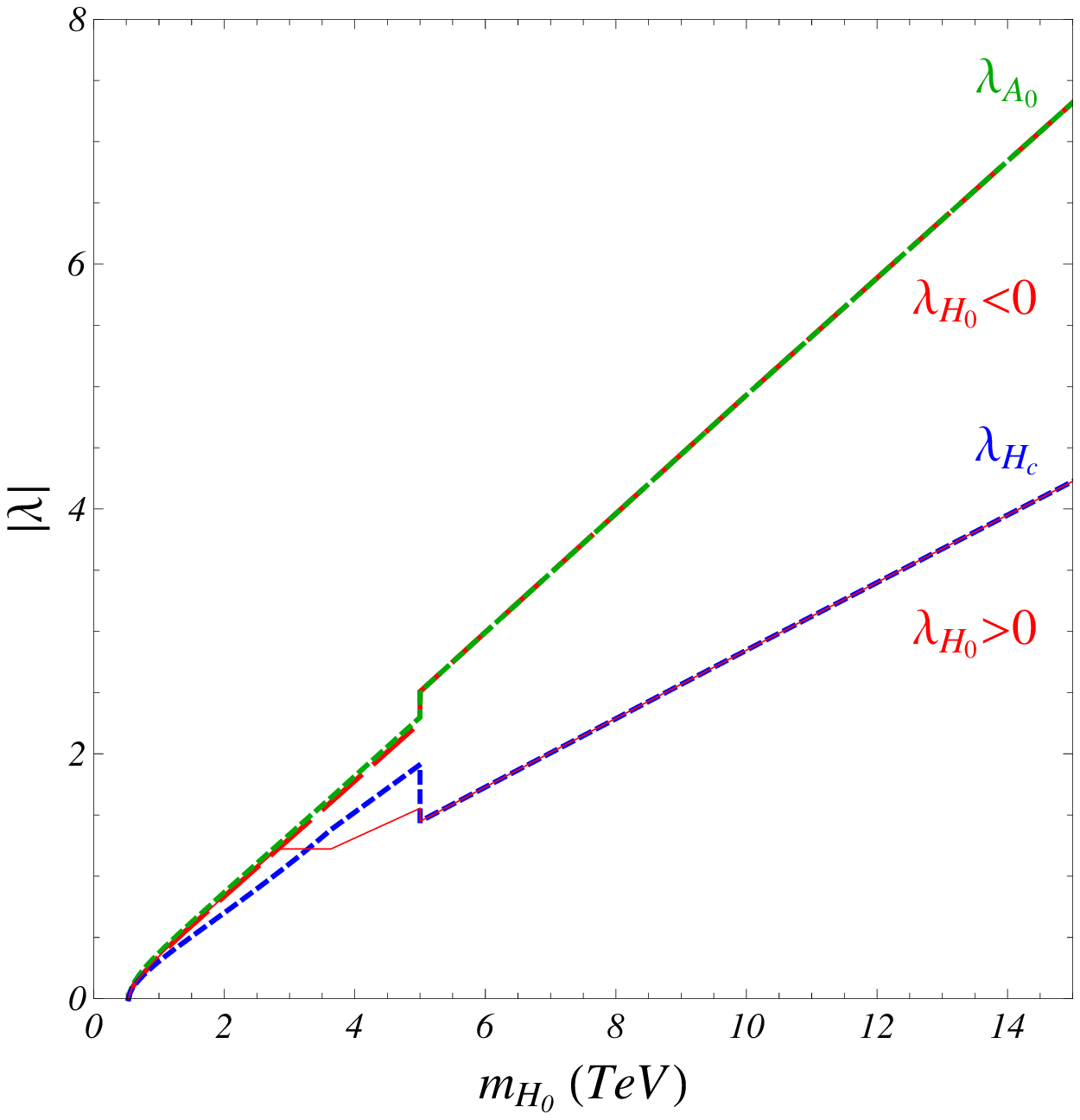}  \quad &
\includegraphics[width=0.4\textwidth]{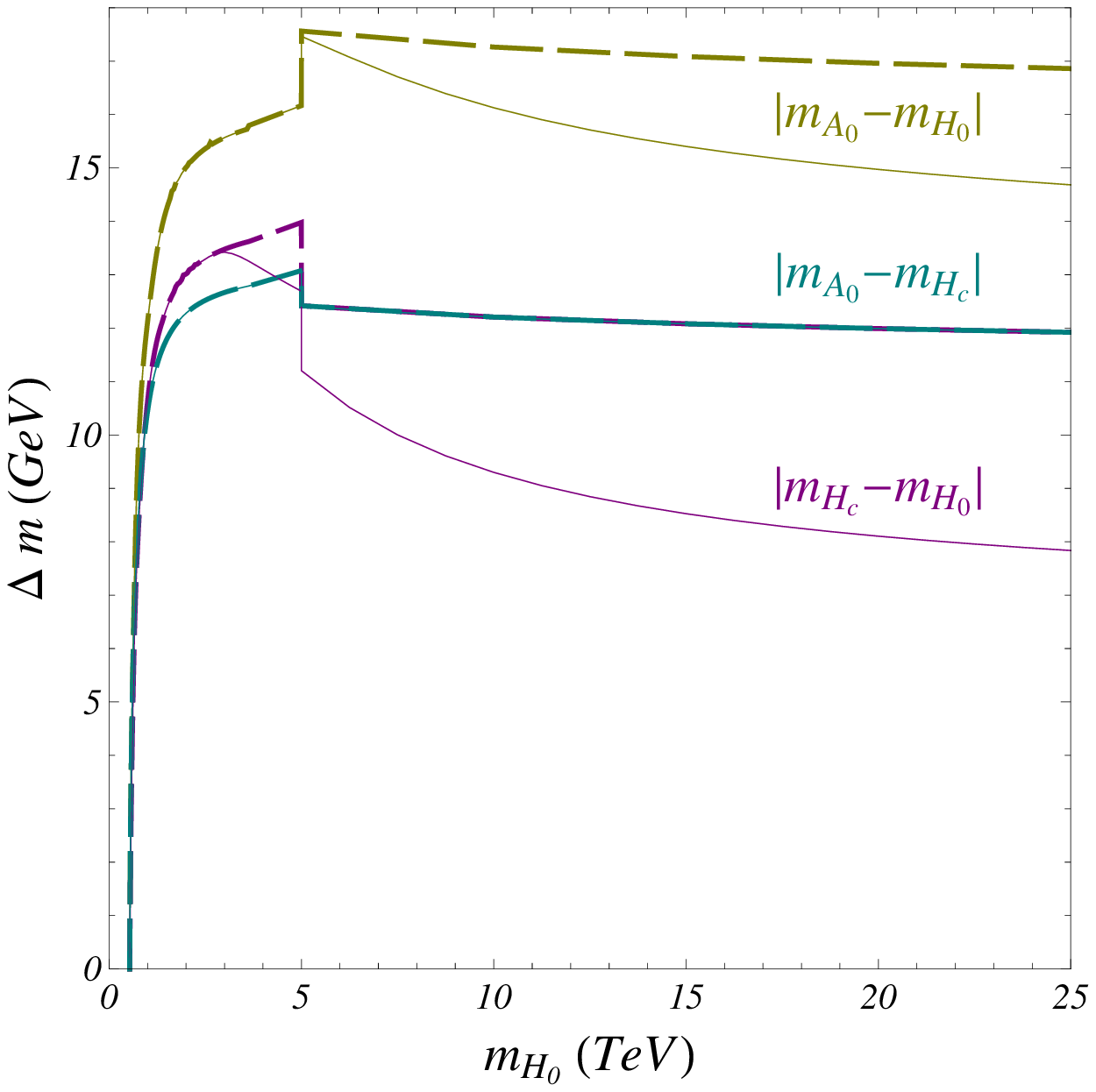}  \quad
\et
\caption{{\it Maximal values of scalar quartic couplings (left panel) and mass splittings (right panel) as a function of the DM mass, constrained by WMAP,
without (dashed lines) and with (thin solid lines) the vacuum stability conditions included.
We assume a Higgs mass $m_h=120$~GeV, and a sharp threshold between the freeze-out in the broken and in the unbroken phases of the SM
at a mass $m_{H_0}=5$~TeV.}}
\label{highlambdamax}
\ec
\efig
For a given DM mass, the values of the scalar quartic couplings corresponding to WMAP lie approximately on an ellipsoid, see Fig.~\ref{lambdacontours}.
\bfig
\bc
\bt{cc}
\includegraphics[width=0.4\textwidth]{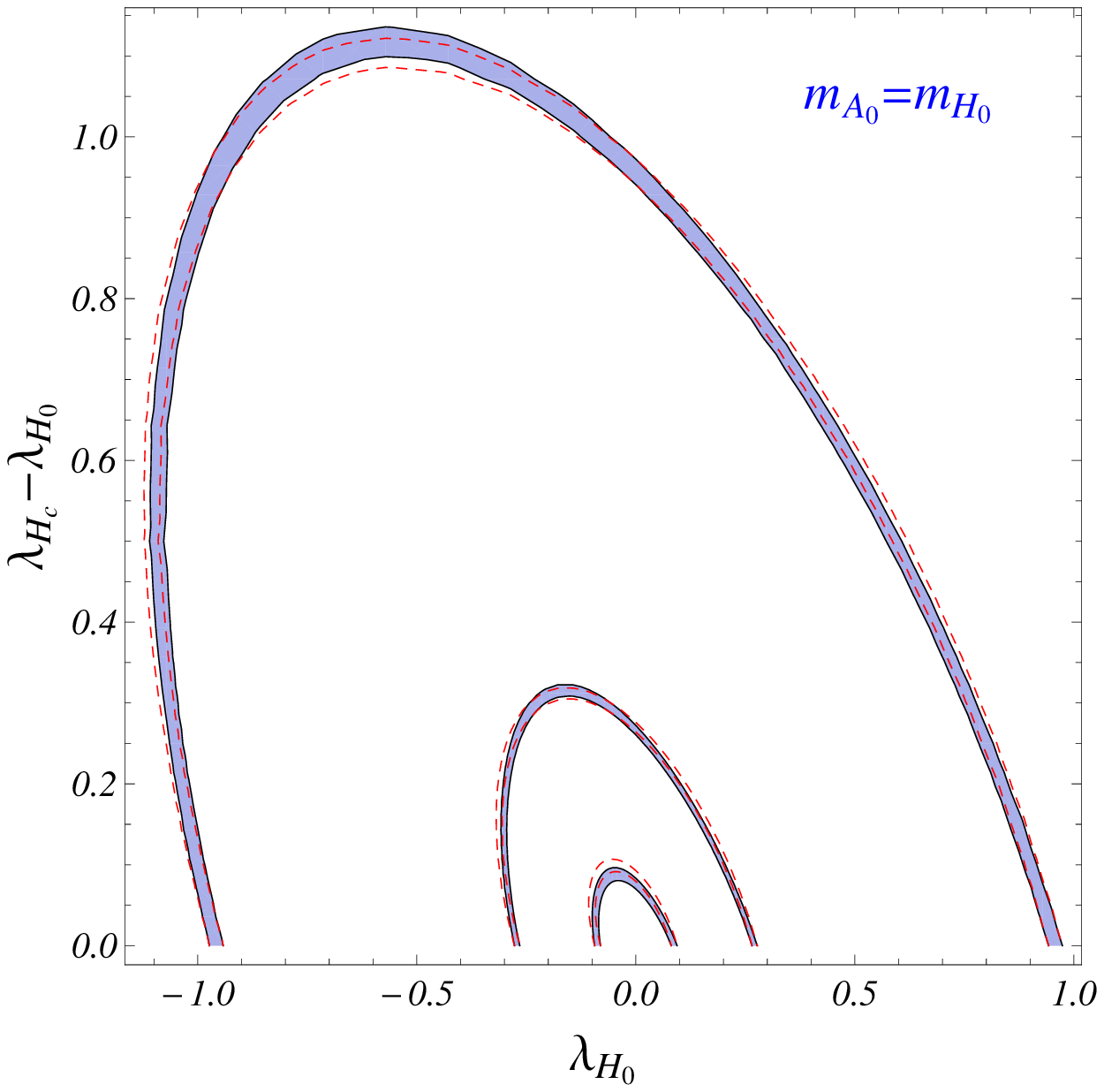}  \quad &
\includegraphics[width=0.4\textwidth]{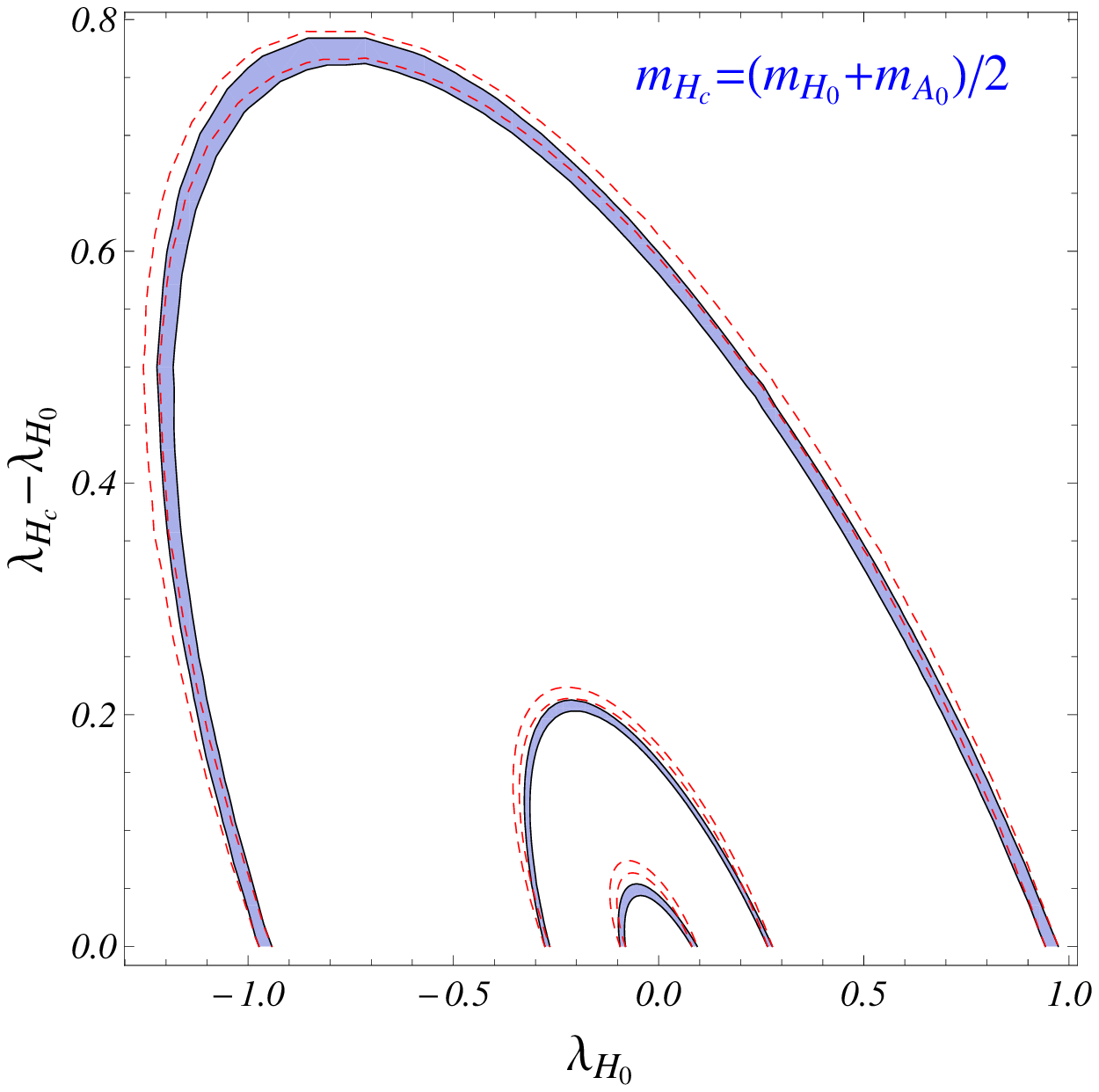}  \quad
\et
\caption{{\it Contours of $\lambda$ for the WMAP value $\Omega_{\rm DM} h^2 = 0.1131
  \pm 0.0034$ for $m_{H_0}=600$ (interior), $1000,\;3000$ (exterior)
  GeV,  
with $m_{A_0}=m_{H_0}$ (left panel) and $m_{H_c}=(m_{H_0}+m_{A_0})/2$ (right panel). 
Red dashed curve corresponds to the approximate ellipsoid.}}
\label{lambdacontours}
\ec
\efig

For higher multiplet models, the annihilation cross-section depends only on one scalar quartic coupling.
Therefore, the WMAP constraint determines this parameter as a function of the DM mass.
As shown in Fig.~\ref{m0lambda}, for very heavy candidates, non-perturbative effects (known as Sommerfeld enhancement
of the annihilation cross-section) due to long range forces become non-negligible.
Their strength increases with the dimension of the multiplet.
\bfig
\bc
\bt{cc}
\includegraphics[width=0.4\textwidth]{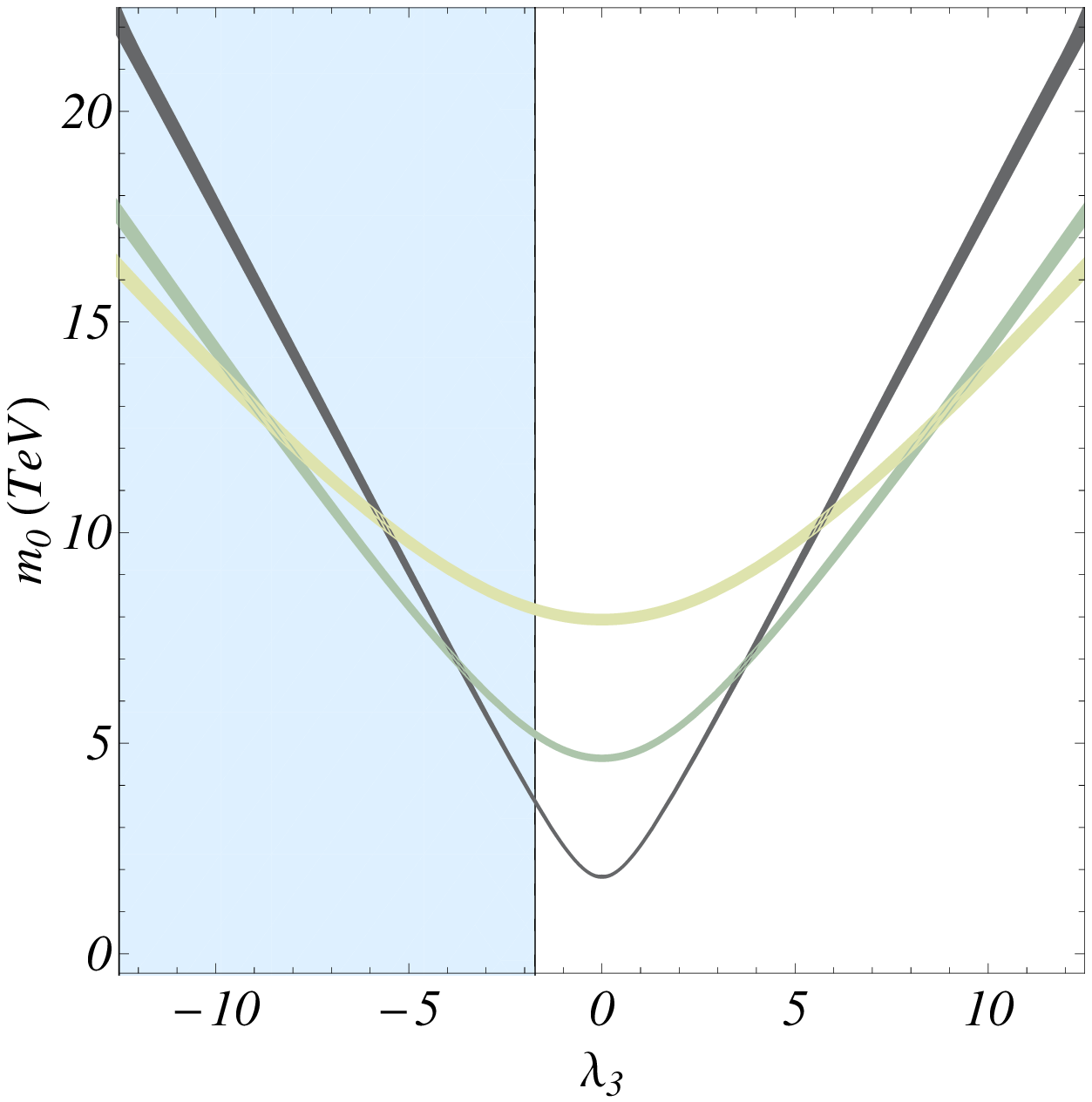}  \quad &
\includegraphics[width=0.4\textwidth]{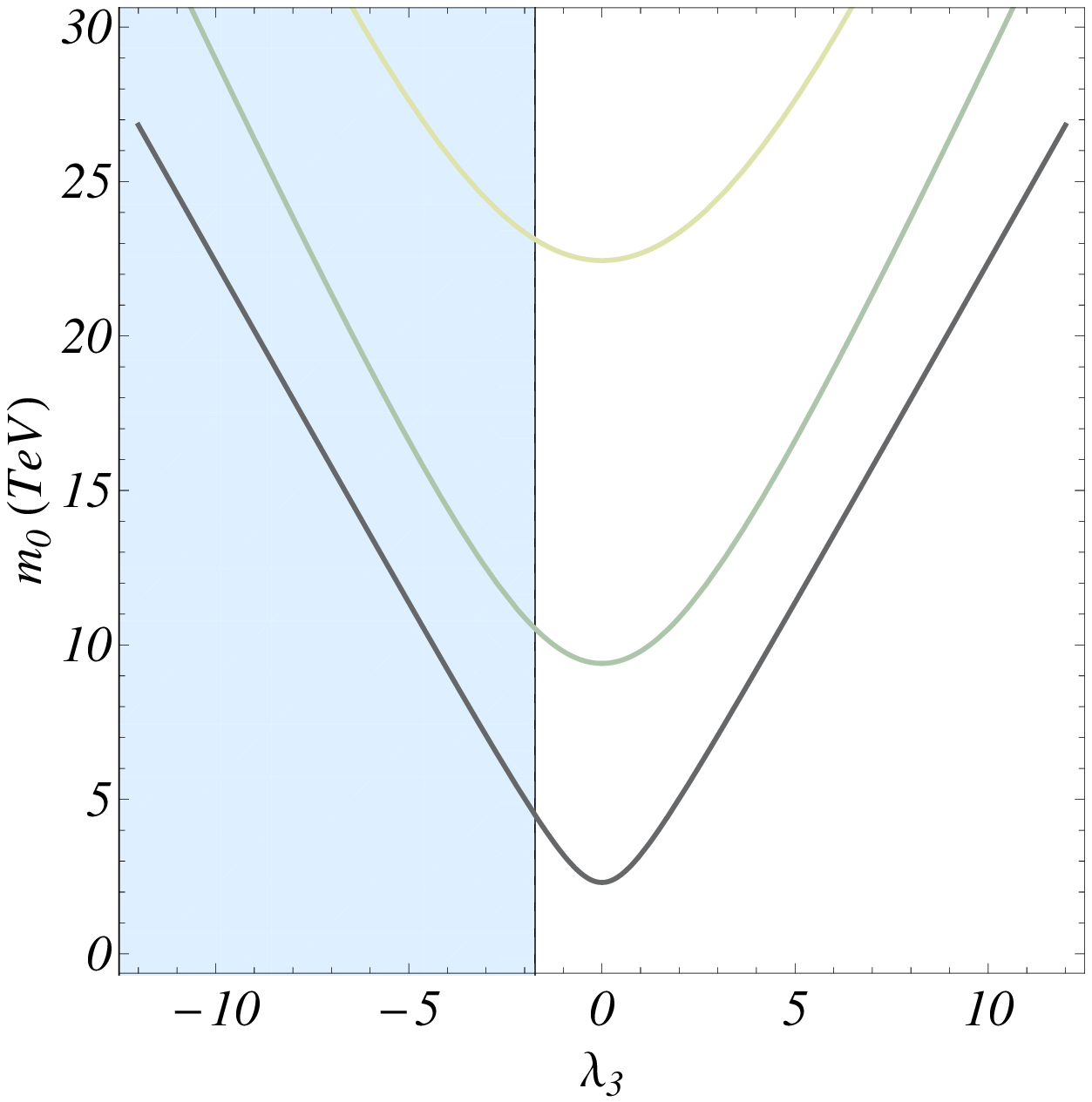}  \quad
\et
\caption{{\it Evolution of the mass of the dark matter candidate as a function of the
coupling $\lambda_3$ for all the higher multiplet models of phenomenological interest,
as constrained by WMAP, without (left panel) or with (right panel) Sommerfeld effect. 
The curves correspond, from top to bottom at $\lambda_3 = 0$, to the real septuplet, the
real quintuplet and the real triplet. The shaded area on the left is excluded by the vacuum stability constraint
(for $m_h=120$~GeV and $\lambda_2^{max}=4\pi$).}}
\label{m0lambda}
\ec
\efig

Therefore, scalar multiplet DM models provide viable candidates with a TeV or multi-Tev mass range.
An upper bound on the DM mass can in principle be derived by imposing that the theory stays perturbative.
The values of the DM mass for $\lambda = 2\pi$ or $\lambda = 4\pi$ (Table~\ref{scatab}) give an indication of the extent 
of the allowed mass range.

\section{Direct \& Indirect detection signals}

Dark matter candidates sensitive to weak interactions give rise to precise and testable predictions in direct and indirect detection experiments.
The pure gauge interactions lead to a minimal cross-section, therefore these candidates cannot be "hidden".
The scalar interactions lead to more freedom and more possibilities compared to a fermionic DM candidate.
Moreover, their strength is constrained by the relic abundance constraint.
As a result, there is also an upper bound on the interaction cross-section at low energy.

These characteristics are illustrated in Fig.~\ref{dddmax} for the direct detection.
\bfig[t]
\bc
\bt{cc}
\includegraphics[width=0.4\textwidth]{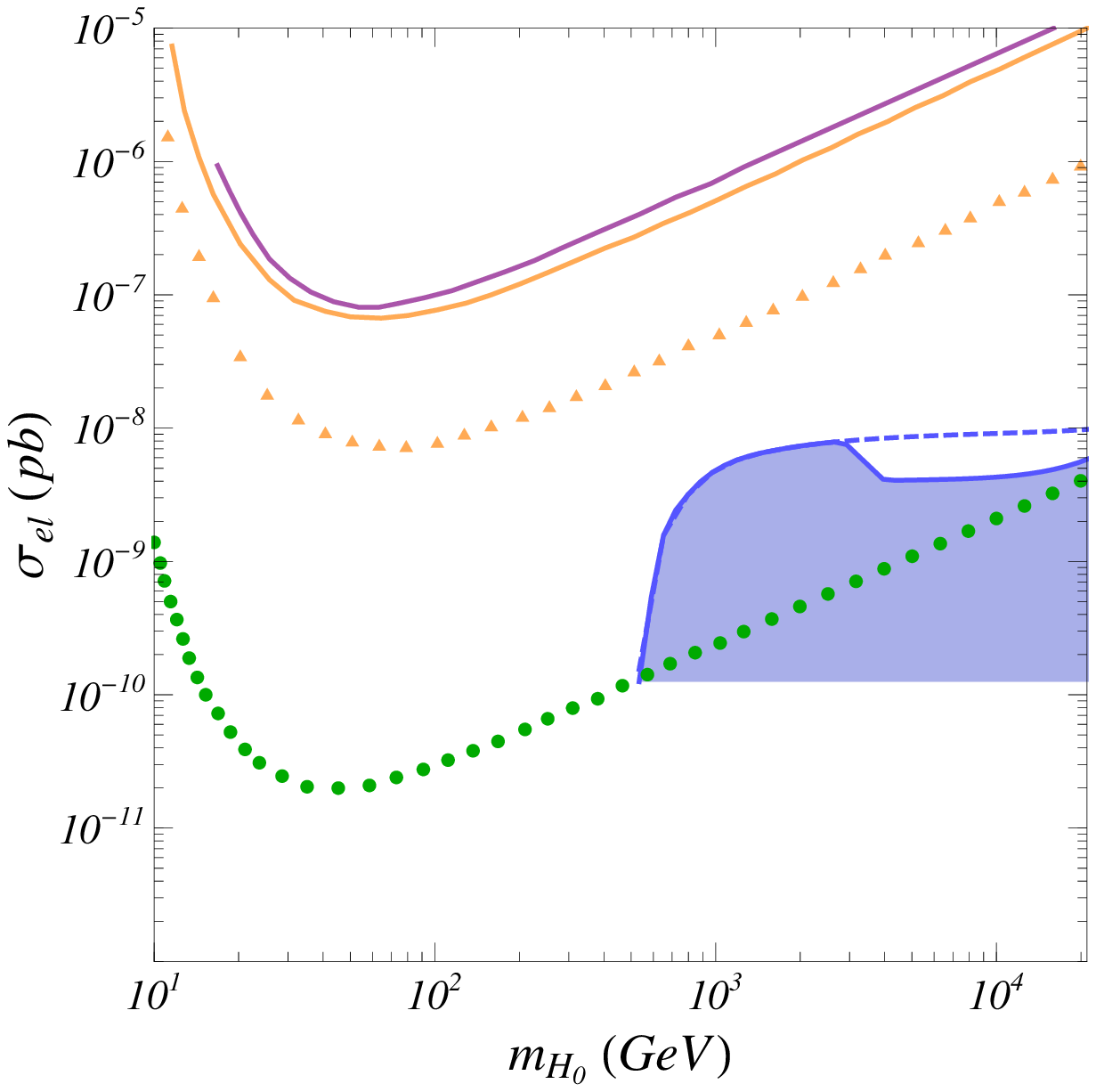} \quad &
\includegraphics[width=0.4\textwidth]{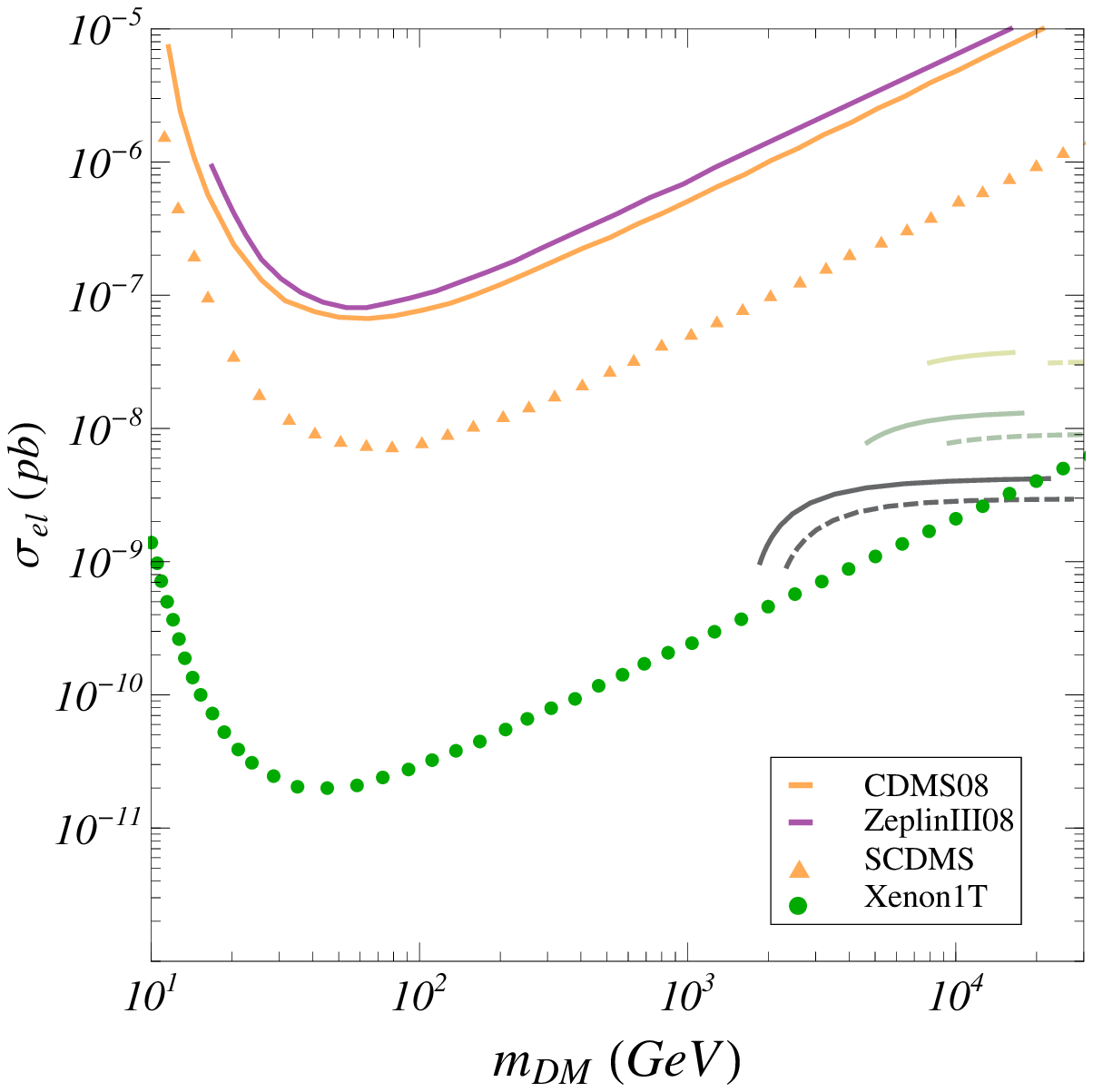} \quad
\et
\caption{{\it Elastic cross-section on nucleon for the inert doublet (left panel) and for higher multiplets (right panel),
compared to experimental limits (CDMS Ge result from 2008~\cite{Ahmed:2008eu}, Zeplin III final result (2008)~\cite{Alner:2007ja}) 
and projected sensitivities at future experiments (Super-CDMS and Xenon 1T)~\cite{DMtools}.
We have assumed $m_h=120$~GeV, a standard Maxwellian DM halo with a local density $\rho_0=0.3~{\rm GeV/cm^3}$.
For the left panel the shaded area gives the allowed range of values. 
Its lower limit corresponds to the pure gauge interaction cross section with zero $\lambda$. 
The upper limit on the elastic cross-section is given by the solid (dashed) blue line when vacuum stability conditions are (not) taken into account.
For the right panel, solid (dashed) curves correspond to the cross-section prediction without (with) Sommerfeld effects.}}
\label{dddmax}
\ec
\efig
In particular, we see that scalar multiplet models will become testable with future direct detection experiments with a ton $\times$ year sensitivity.

For indirect searches, the most promising  signal is the observation of $\gamma$ rays from the galactic center,
with a possible complementarity with high energy neutrinos. 
The signal will be observable by the FERMI-LAT satellite if the galactic DM halo is cuspy enough, and if the DM mass is not too heavy
(The number density of DM particles which controls the annihilation rate is obviously inversely proportional to the DM mass).
Moreover, annihilation signals can benefit from a possible Sommerfeld enhancement.
As scalar DM candidates are viable for a continuous range of mass, for some values, a resonance phenomenon can occur.
The absolute boost factor due to particle physics enhancements is however limited by constraints from existing $\gamma$ ray
measurements like the EGRET data~\cite{cirellipanci}.

The complementairity of different searches is illustrated by the fact that the main primary annihilation 
channels of scalar multiplet DM candidates are $W^+ W^-$, $Z Z$ and $h h$.
Therefore, the production rates of photons, neutrinos and charged cosmic rays are determined by the subsequent decays and hadronization processus involved
by these particles. In particular, it appears that the charged cosmic ray fluxes (antiprotons and positrons) are several orders of magnitude
below the observed background. The recent positron excesses claimed by both the PAMELA and the ATIC experiments cannot be explained in this context,
unless an important boost factor is applied. Such a boost factor would however again lead to a gamma ray flux in excess of the EGRET data for most of the parameter
space.



\end{document}